\documentclass[twocolumn,aps,pra,showpacs]{revtex4-1}

\usepackage{amssymb}
\usepackage{mathrsfs}
\usepackage{lipsum}
\usepackage{graphicx}
\usepackage[caption=false]{subfig}
\usepackage{mathtools}
\usepackage{epstopdf}
\usepackage{xcolor}
\usepackage[colorlinks, linkcolor=red, anchorcolor=green, citecolor=blue, urlcolor=violet]{hyperref}
\DeclareGraphicsExtensions{.pdf,.jpg,.png,.eps}
\usepackage[toc,page,title,titletoc,header]{appendix}

\begin{document}
	
\title{Sub-Heisenberg-limited nonlinear phase estimation: Parity measurement approaches the quantum Cram\'er-Rao bound}

\author{Jian-Dong Zhang}
\affiliation{Department of Physics, Harbin Institute of Technology, Harbin, 150001, China}
\author{Zi-Jing Zhang}
\email[]{zhangzijing@hit.edu.cn}
\affiliation{Department of Physics, Harbin Institute of Technology, Harbin, 150001, China}
\author{Long-Zhu Cen}
\affiliation{Department of Physics, Harbin Institute of Technology, Harbin, 150001, China}
\author{Jun-Yan Hu}
\affiliation{Department of Physics, Harbin Institute of Technology, Harbin, 150001, China}
\author{Yuan Zhao}
\email[]{zhaoyuan@hit.edu.cn}
\affiliation{Department of Physics, Harbin Institute of Technology, Harbin, 150001, China}

\date{\today}
	
\begin{abstract}
Quantum-enhanced phase estimation paves the way to ultra-precision sensing and is of great realistic significance.
In this paper we investigate theoretically the estimation of a second-order nonlinear phase shift using a coherent state and parity measurement. 
A numerical expression is derived, the resolution and sensitivity of parity signal are contrasted to linear phase estimation protocol, and the signal visibility is analyzed.
Additionally, by virtue of phase-averaging approach to eliminate any hidden resources, we make an attempt at unveiling the low-down on the fundamental sensitivity limit from quantum Fisher information.
Finally, the effects of several realistic scenarios on the resolution and the sensitivity are studied, including photon loss, imperfect detector, and those which are a combination thereof.		
\end{abstract}

\pacs{42.50.Dv, 42.50.Ex, 03.67.-a}
	

\maketitle

\section{Introduction}
Optical interferometers are one of the most fundamental tools to offer insight into slight variations on numerous physical quantities, phase shifts \cite{Nagata726,PhysRevLett.119.080502,Schafermeier:18,PhysRevLett.111.033603}, polarized rotations \cite{PhysRevA.96.053846,Zhang:17}, angular displacements \cite{D2013Photonic,PhysRevLett.112.200401,PhysRevA.83.053829}, to name a few.
As two performances most concerned for interferometric measurements, the resolution and the sensitivity of the conventional interferometers are limited by the Rayleigh diffraction and shot-noise limit, respectively.
Various quantum technologies are predicted to be able to break through these no-go areas, and the corresponding developments$\---$quantum-enhanced interferometers$\---$also arouse great research interests.
Over the past few years, estimating linear phase shifts with exotic quantum resources has become an indispensable component of quantum-enhanced interferometric metrology, accordingly, novel results are popping out from time to time.
A great deal of strategies that can improve the resolution are presented to beat the Rayleigh diffraction, such as parity measurement \cite{Cohen:14,Gao:10,PhysRevA.54.R4649}, Z measurement \cite{Cohen:14,Zhang2018Optimal}, and homodyne measurement \cite{PhysRevLett.111.033603,PhysRevA.90.013807}.
As to sensitivity, the Heisenberg-limited and even sub-Heisenberg-limited sensitivities have been shown with utilizing twin Fock \cite{PhysRevA.68.023810}, N00N \cite{PhysRevLett.112.103604}, entangled coherent \cite{PhysRevLett.107.083601}, and two-mode squeezed vacuum states \cite{PhysRevLett.104.103602}.

Although these exotic quantum states provide eximious performances, an insurmountable fact that has to be acknowledged is the preparing difficulty for large mean photon number.
In the previous experiments, $\bar N=2.6$ and $\bar N=4$ corresponding to coherent superposition states \cite{Ourjoumtsev2007Generation} and squeezed vacuum \cite{PhysRevLett.100.033602} have been generated, the experimental preparations for biphoton and triphoton Fock states \cite{PhysRevLett.96.213601,Cooper:13} have also been reported.
In the scenarios when the high-intensity input is allowable, as a result of restrictions on the photon number which can be produced, the performances of quantum states are even inferior to those of a coherent state, i.e., the available photon numbers of quantum states downplay their quantum advantages.
Hence, coherent-state-based researches are still a remarkable topic in interferometric optical metrology \cite{PhysRevLett.100.073601,PhysRevLett.111.173601,PhysRevA.81.033819}.

Besides the above, in recent years, employing nonlinear elements for optical interferometers has also been studied extensively.
A typical achievement is is a SU(1,1) interferometer, also called nonlinear interferometer, which replaces two beam splitters in SU(2) one by two optical parametric amplifiers \cite{PhysRevA.33.4033,Chekhova:16}.
On the other hand, nonlinear phase processes have gradually received the attentions, and have demonstrated many significant physical phenomena, e.g., preparation for a Schr\"odinger cat state, or rather, a Schr\"odinger kitten state \cite{PhysRevA.45.6811,PhysRevLett.107.083601}.
However, the studies on nonlinear phase estimation \cite{PhysRevA.34.3974,PhysRevA.66.013804,PhysRevA.90.063838,PhysRevA.88.013817,PhysRevA.86.043828,PhysRevLett.105.010403} are not as deep-going and systematic as those on linear phase estimation.
Among these researches, most only work on the sensitivity limit based upon direct calculation of quantum Fisher information \cite{PhysRevLett.72.3439}, i.e., a special measurement saturated with the limit remains to be provided \cite{PhysRevA.90.063838,PhysRevA.88.013817,PhysRevA.86.043828,PhysRevLett.105.010403}.
Meanwhile, these discussions may loose the tightness of quantum Fisher information due to the ill-considered selection of evolution operator in calculation, consequently, the delightful calculation result is likely to be a pitfall, which cannot be realized at all via concrete input state and measurement.
	
In this paper, we focus on discussing the estimation of a nonlinear phase shift with utilizing a coherent state and parity measurement. 
For avoiding the aforementioned problem and addressing such underlying hazard, we take advantage of the phase-averaging approach \cite{PhysRevA.85.011801,PhysRevA.96.052118} to ascertain an authentic sensitivity limit for our protocol.
To our knowledge, it is the first ever implementation for nonlinear phase estimation.
This approach rules out the virtual advantages brought by hidden resources in quantum Fisher information, and unveils the low-down on the fundamental limit which can be achieved from exact input and measurement.
Such approach can also bypass the troublesome choice for evolution operator form in calculation.
	
The remainder of this paper is organized as follows.	
Section \ref{s2} introduces the framework and working principle for our protocol.
In Sec. \ref{s3}, we show the resolution and the sensitivity of the signal with parity measurement, and compare them with linear phase estimation protocol.
We explore the fundamental sensitivity limit by invoking phase-averaging approach and quantum Fisher information in Sec. \ref{s4}.
In Sec. \ref{s5}, the effects of several realistic scenarios on the resolution and the sensitivity are discussed.
Finally, we summarize our work with a conclusion in Sec. \ref{s6}.

\section{Nonlinear phase estimation protocol}
\label{s2}
\begin{figure}[htbp]
\centering\includegraphics[width=8cm]{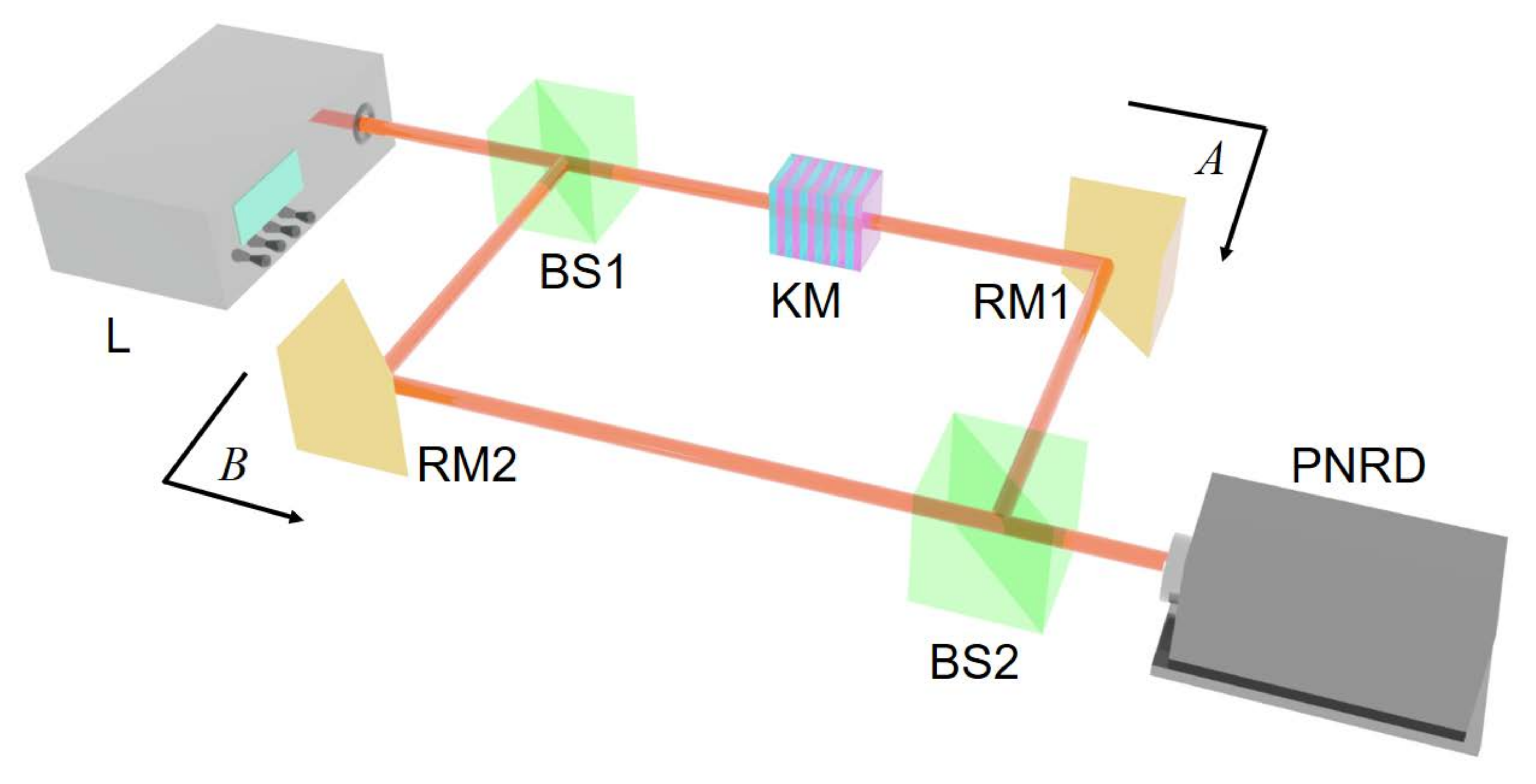}
\caption{Schematic diagram of the measuring protocol for nonlinear phase estimation. L, laser; BS, beam splitter; RM, reflection mirror; KM, Kerr medium; PNRD, photon-number-resolving detector.}
\label{f1}
\end{figure}

We start off with the measuring protocol describing the nonlinear phase estimation, as illustrated in Fig. \ref{f1}.
The balanced Mach-Zehnder interferometer considered here is formed by two fifty-fifty beam splitters and a nonlinear medium, the two input ports are fed by a coherent state $\left| \alpha  \right\rangle$ generated from laser and a vacuum state $\left| 0  \right\rangle$, respectively. 
Where the clockwise path in the interferometer is labeled as spatial path $A$, and the counterclockwise one is marked as $B$.
The corresponding input state can be written as $\left| {\alpha ,0} \right\rangle  \equiv {\left| \alpha  \right\rangle _A}{\left| 0 \right\rangle _B}$.  
After the first beam splitter, the state is divided into two coherent states, which have the direct product representation ${\left| {{\alpha  \mathord{\left/
				{\vphantom {\alpha  {\sqrt 2 }}} \right.
				\kern-\nulldelimiterspace} {\sqrt 2 }}} \right\rangle _A}{\left| {{\alpha  \mathord{\left/
				{\vphantom {\alpha  {\sqrt 2 }}} \right.
				\kern-\nulldelimiterspace} {\sqrt 2 }}} \right\rangle _B}$.
	
For the nonlinear phase channel, a Kerr medium is embedded into path $A$ for introducing a nonlinear phase shift $\varphi$, the parameter we would like to estimate.
The generalized formulation toward nonlinear phase operator ${\hat U_k}\left( \varphi  \right) $ is given by
\begin{equation}
{\hat U_k}\left( \varphi  \right) = \exp \left[ {i\varphi {{\left( {{{\hat a}^\dag }\hat a} \right)}^k}} \right],
\label{e1}
\end{equation}
where ${{\hat a}^\dag }$ (${\hat a}$) stands for a creation (annihilation) operator in path $A$, the exponent $k$ represents the order of the nonlinearity, and the nonlinear effect in this paper only refers to the case of $k=2$.
The nonlinear phase follows $\varphi  = \chi t$, where $\chi$ is proportional to the third-order nonlinear susceptibility $\chi^{(3)}$, and $t$ is the time for the light to cross the medium \cite{PhysRevA.66.013804}.
When the state passes the nonlinear phase channel, the resultant state arrives at
\begin{align}
\nonumber  \left| \psi  \right\rangle  &= {{\hat U}_2}\left( \varphi  \right)\left| {\frac{\alpha }{{\sqrt 2 }},\frac{\alpha }{{\sqrt 2 }}} \right\rangle  \\ 
	& = \exp \left( { - \frac{N}{2}} \right) {\sum\limits_{m,n = 0}^\infty  {\frac{{{{\left( {{{\alpha {e^{in\varphi }}} \mathord{\left/
										{\vphantom {{\alpha {e^{in\varphi }}} {\sqrt 2 }}} \right.
										\kern-\nulldelimiterspace} {\sqrt 2 }}} \right)}^n}}}{{\sqrt {n!} }}} } \frac{{{{\left( {{\alpha  \mathord{\left/
								{\vphantom {\alpha  {\sqrt 2 }}} \right.
								\kern-\nulldelimiterspace} {\sqrt 2 }}} \right)}^m}}}{{\sqrt {m!} }}\left| {n,m} \right\rangle 
\label{e2}
\end{align}
with twin Fock basis, where $N={{\left| \alpha  \right|}^2}$ denotes the mean photon number in the input. 
Finally, this state is incident on the second beam splitter, and the measurement and estimation is performed onto the output.
Especially, for $\varphi=\pi/2$, after tracing over the path $B$, the reduced state in Eq. (\ref{e2}) is a superposition of two coherent states, and the entire output evolves into an entangled coherent states.

\section{Resolution and sensitivity}
\label{s3}
We consider the parity measurement that monitors whether the photon number in a given output port is even or odd.
It is first proposed by Bollinger $et$ $al.$ for trapped ions \cite{PhysRevA.54.R4649}, subsequently, Gerry applies it to the field of optical metrology \cite{PhysRevA.61.043811,PhysRevA.72.053818}, and the work of Plick $et$ $al.$ reduces the stringent requirements of this strategy towards detectors \cite{1367-2630-12-11-113025}. 
As an excellent binary strategy, parity measurement plays a nontrivial role in linear phase estimation, and has been proved to be the optimal measurement strategy for a number of protocols \cite{PhysRevLett.104.103602,PhysRevA.87.043833,1367-2630-13-8-083026}.
Take output port $B$ as an instance, the parity operator reads as $\hat \Pi = \exp \left( {i\pi {{\hat n}_B}} \right)$.

Note that performing parity measurement onto either of two outputs is equivalent  to the implementation of projective measurement to the state before the second beam splitter \cite{Gao:10}, thus we have

\begin{align}
\nonumber \left\langle {\hat \Pi } \right\rangle  &={\rm Tr}\left\{ {\hat U_{\rm BS}}\left| \psi  \right\rangle\left\langle \psi  \right|\hat U_{\rm BS}^\dag \left[\hat I_A \otimes \exp \left( {i\pi {{\hat n}_B}} \right)\right]  \right\}\\
&= \left\langle \psi  \right|{\hat \mu _{AB}}\left| \psi  \right\rangle. 
\label{e3}
\end{align}
Where $\hat I$ is the identity operator, the operator for the fifty-fifty beam splitter can be formulated as
\begin{equation}
{\hat U_{\rm {BS}}} = \exp \left[ {i\frac{\pi }{4}\left( {{{\hat a}^\dag }\hat b + {{\hat b}^\dag }\hat a} \right)} \right], 
\label{e4}
\end{equation}
and the projective operator ${\hat \mu _{AB}}$ has the following representation in twin Fock basis,
\begin{equation}
{\hat \mu _{AB}} = \sum\limits_{n' = 0}^\infty  {\sum\limits_{m' = 0}^\infty  {\left| {n',m'} \right\rangle \left\langle {m',n'} \right|} }. 
\label{e5}
\end{equation}
Consequently, the expectation value of parity operator is rewritten as
\begin{equation}
\left\langle {\hat \Pi } \right\rangle  = {e^{ - N}}\sum\limits_{n = 0}^\infty  {\sum\limits_{m = 0}^\infty  {\frac{{{{\left( {{{N{e^{in\varphi }}} \mathord{\left/
										{\vphantom {{N{e^{in\varphi }}} 2}} \right.
										\kern-\nulldelimiterspace} 2}} \right)}^n}}}{{n!}}} } \frac{{{{\left( {{{N{e^{ - im\varphi }}} \mathord{\left/
								{\vphantom {{N{e^{ - im\varphi }}} 2}} \right.
								\kern-\nulldelimiterspace} 2}} \right)}^m}}}{{m!}}.
\label{e6}
\end{equation}
	
This double series sum has no ready-made formula to follow. 
In order to simplify the calculation, we adopt numerical approach by terminating the sum at a sufficiently large total probability.
By analyzing available quantum Fisher information rooting in the $\left( n'+m'\right) $-photon weight in the input \cite{PhysRevA.95.023824}, we configure the truncated upper limit adhering to 
\begin{equation}
\max \left[ {n'} \right] = \max \left[ {m'} \right] \geqslant 5N,
\label{e7}
\end{equation}
this condition ensures that the signal has a terrific fidelity.
	
\begin{figure}[htbp]
\centering\includegraphics[width=7cm]{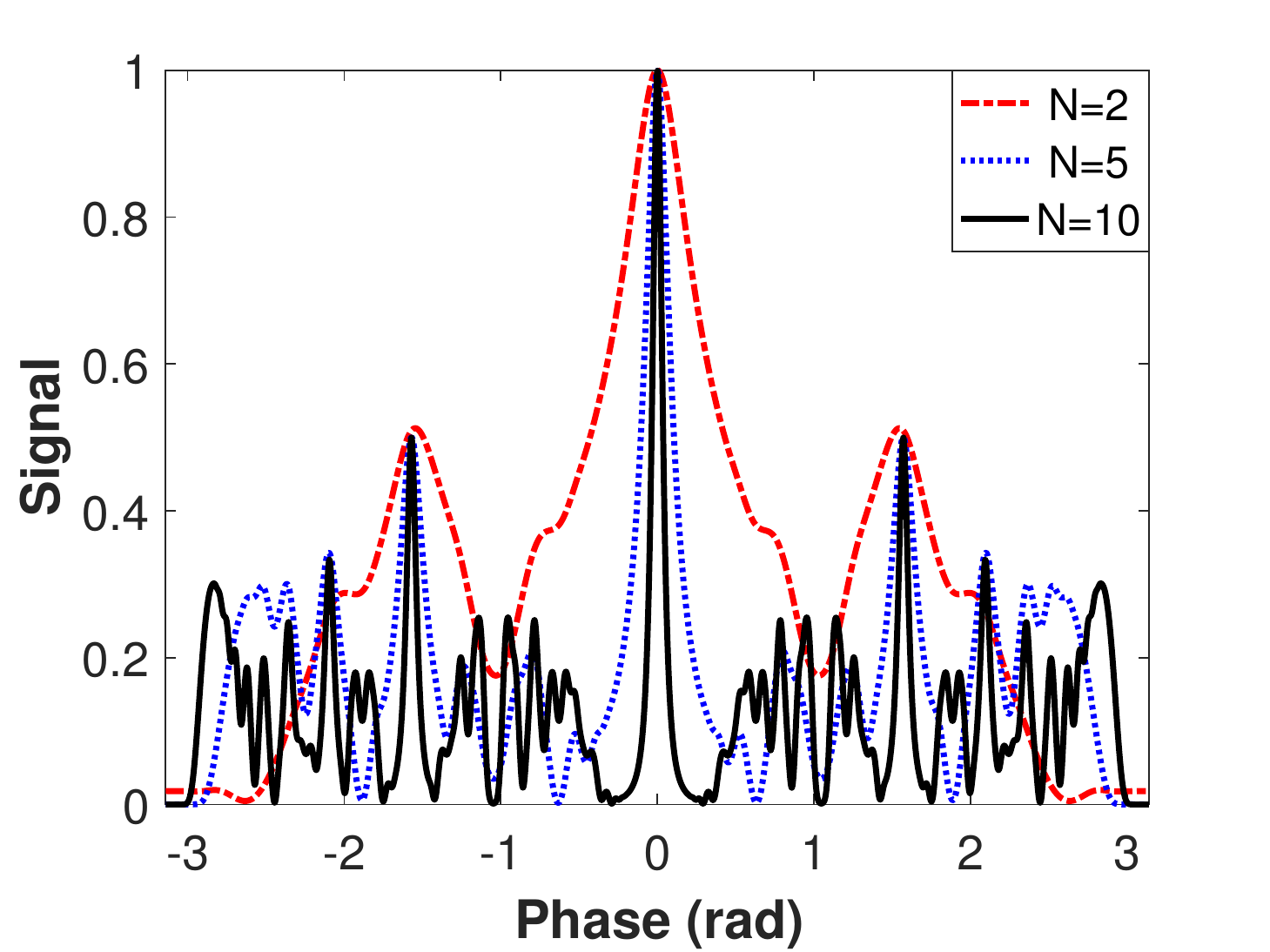}
\caption{The signal with parity measurement as a function of the phase, where red dash-dotted, blue dotted and black solid lines correspond to the scenarios of $N=2$, 5 and 10, respectively.}
\label{f2}
\end{figure}

To intuitively observe the behaviors of the output, in Fig. \ref{f2} we plot the parity signal as a function of the nonlinear phase with different mean photon numbers.
Unlike the scenario of linear phase estimation, there exists a bunch of miscellaneous peaks around the main peak ($\varphi=0$), and the number of peaks is increasing as the increase of mean photon number. 
On the other hand, the maxima of these peaks are less than 1$\---$the maximum of the signal$\---$whereupon we take no notice of them, and only focus on the main peak.
As the super-resolution criterion, the full width at half maximum (FWHM) of the main peak gets narrow when increasing mean photon number, i.e., the signal's super-resolved characteristic becomes more obvious.
Moreover, the signal possesses a 100\% visibility \cite{PhysRevLett.121.026802,doi:10.1080/00107510802091298}, in that the minimum sits at 0.
The definition of visibility refers to
\begin{equation}
V = \frac{{{{\left\langle {\hat \Pi } \right\rangle }_{\max }} - {{\left\langle {\hat \Pi } \right\rangle }_{\min }}}}{{{{\left\langle {\hat \Pi } \right\rangle }_{\max }} + {{\left\langle {\hat \Pi } \right\rangle }_{\min }}}}.
\label{e8}
\end{equation}

In addition, the signal takes on the value of 1 for $\varphi=0$, this means that the parity is always even in port $B$.
At this point the whole interferometer can be regarded as an identity operator, accordingly, the output in port $B$ remains the vacuum state.

\begin{figure}[htbp]
	\centering\includegraphics[width=7cm]{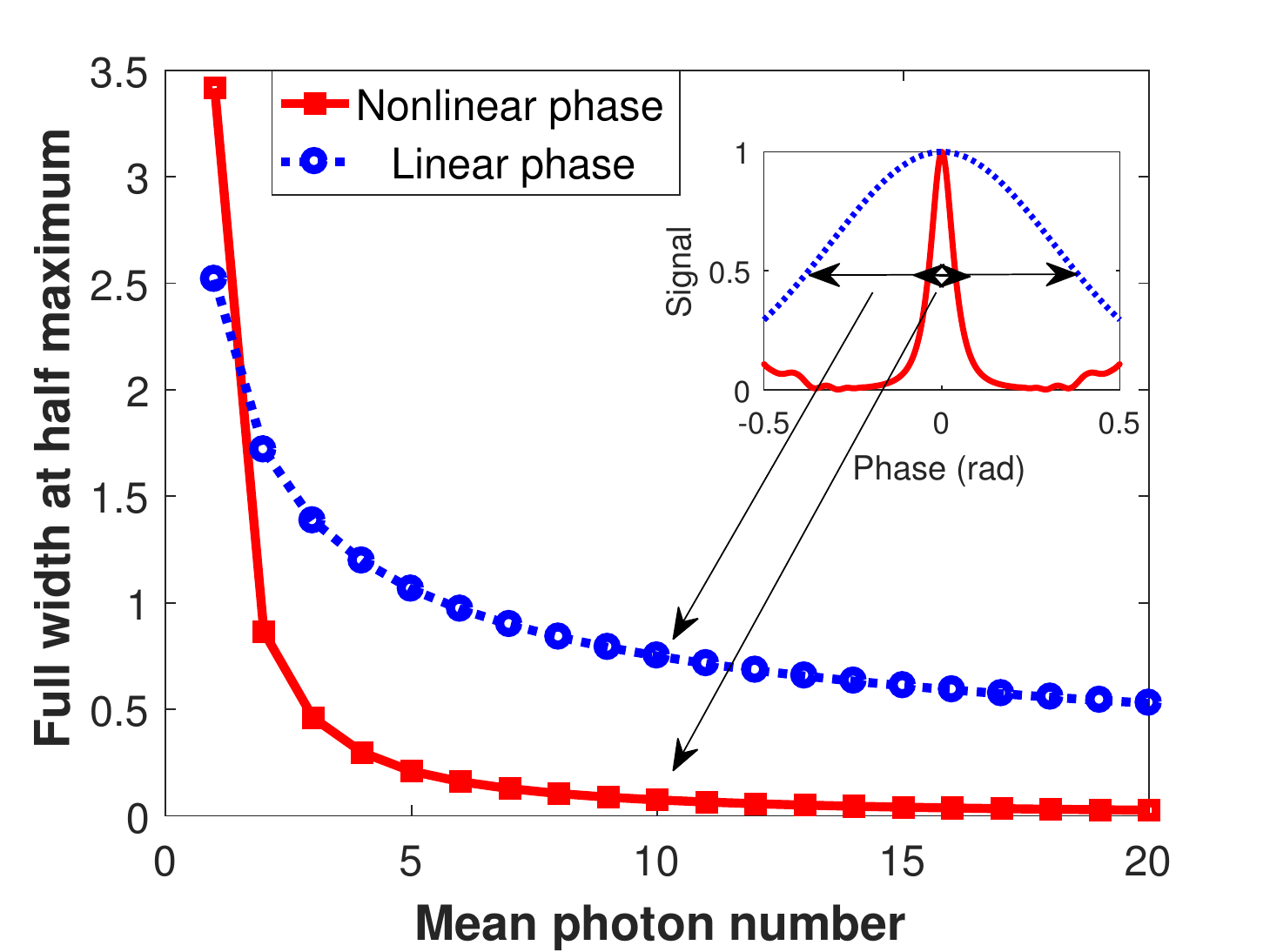}
	\centering\includegraphics[width=7cm]{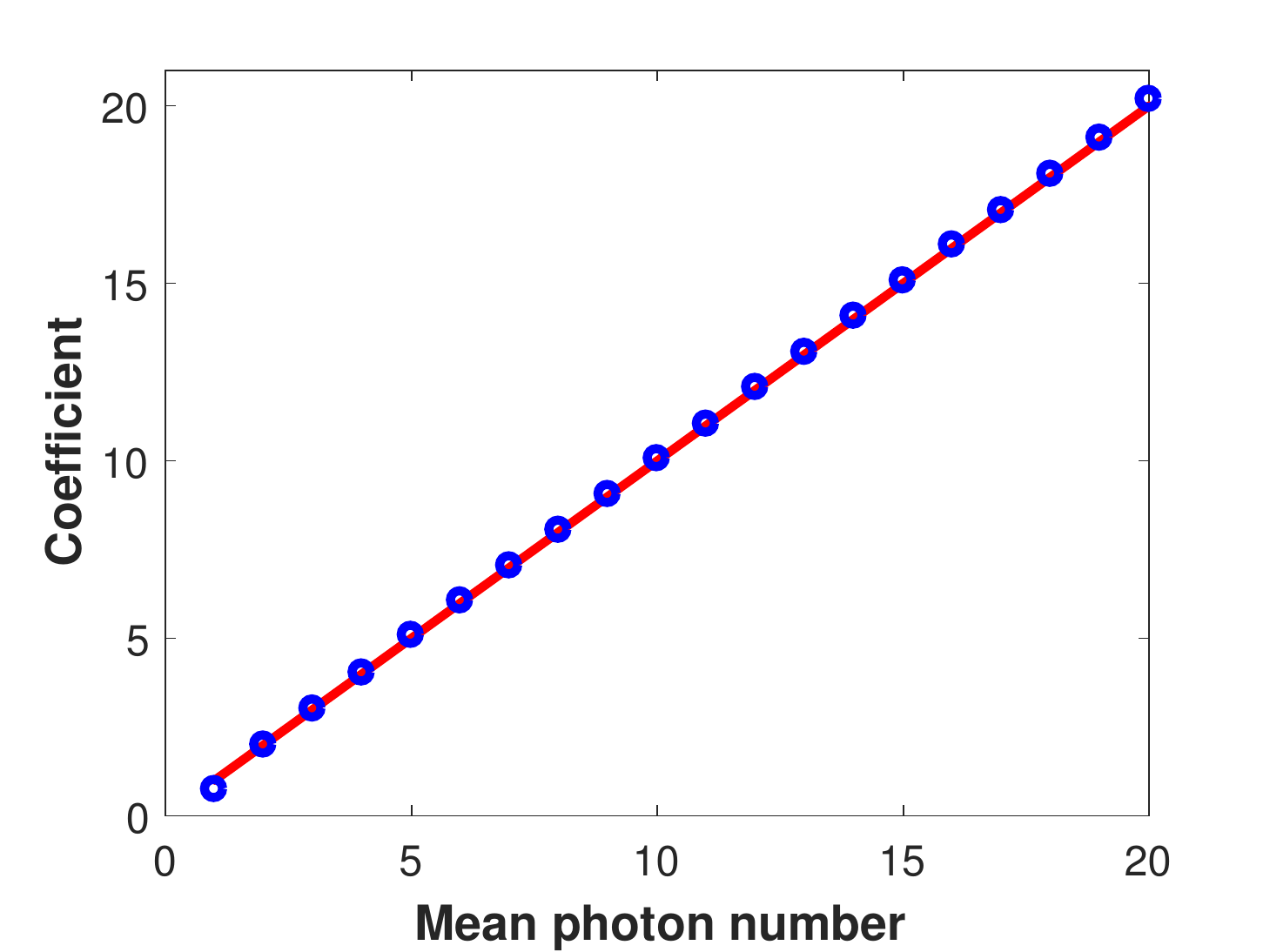}
	\caption{(a) The FWHMs of nonlinear and linear phase protocols as functions of the mean photon number, where the blue dotted and red solid lines represent linear and nonlinear phase protocols, respectively. 
		As an auxiliary example, the subgraph shows the details of linear and nonlinear signals with $N=10$.
		(b) The coefficient, the ratio of FWHM of linear phase signal to that of nonlinear one, as a function of the mean photon number, where the red solid line is a reference line with $C=N$.}
	\label{f3}
\end{figure}

In order to quantify the super-resolved characteristic, in Fig. \ref{f3}(a) we exhibit the FWHMs of estimation signals for linear and nonlinear phase protocols.
The result indicates that the FWHM of nonlinear protocol stands in stark contrast with that of linear one and is narrower than the latter for $N \geqslant 2$, that is, the super-resolution of nonlinear protocol is more apparent.
Figure \ref{f3}(b) clearly illustrates the difference between the two scenarios.
Hereon we define a coefficient to appraise the nonlinear advantage compared to the linear protocol, the coefficient takes the form:
\begin{equation}
C = \frac{{{\mathop{\rm FWHM}\nolimits} \left[ {{{\left\langle {\hat \Pi } \right\rangle }_{k = 1}}} \right]}}{{{\mathop{\rm FWHM}\nolimits} \left[ {{{\left\langle {\hat \Pi } \right\rangle }_{k = 2}}} \right]}}.
\label{e9}
\end{equation}

According to this definition, we plot the coefficient $C$ as a function of the mean photon number.
As shown in Fig. \ref{f3}(b), the coefficient  increases monotonically as the increase of photon number, and we provide a reference line with $C=N$.
This means that, with the same photon number as input, the super-resolution of nonlinear protocol surpasses that of linear one by a factor of $N$ or so.
The $N$-based boosted resolution may come from the phase shift $\exp\left( in^2\varphi \right) $ for $n$-photon state, while in linear phase protocol this shift is $\exp\left( in\varphi \right) $.

\begin{figure}[htbp]
	\centering\includegraphics[width=7cm]{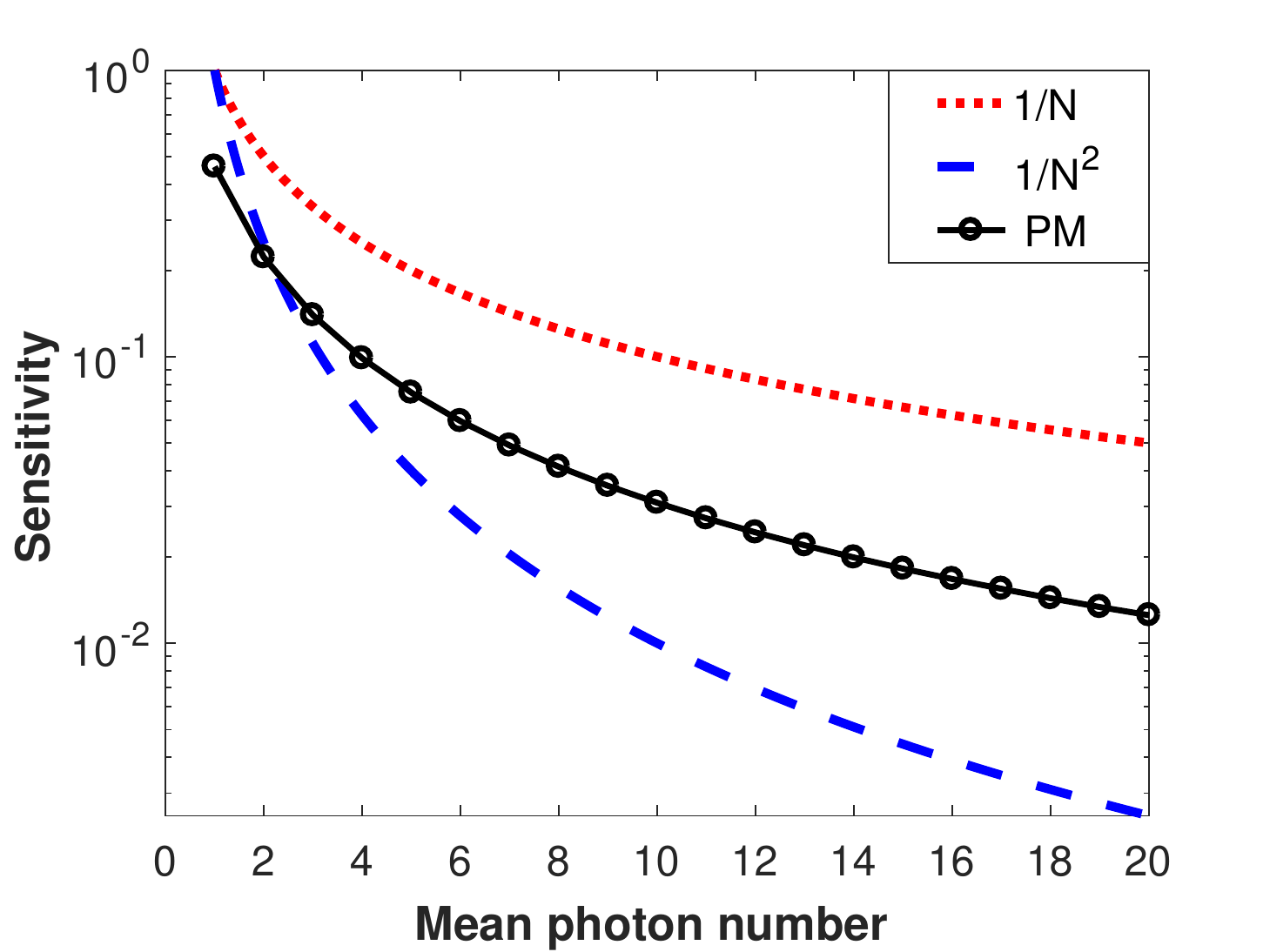}
	\caption{The optimal sensitivity with parity measurement as a function of the mean photon number, where red dotted and blue dashed lines scale with $N^{-1}$ and $N^{-2}$, respectively.}
	\label{f4}
\end{figure}

Now we turn to the discussion of the sensitivity.
Using the classical Fisher information ${\cal F_{\rm c}}$ \cite{PhysRevA.62.012107}, we can calculate the phase sensitivity
\begin{equation}
\delta \varphi  = \frac{1}{{\sqrt {\cal F_{\rm c}} }} = {\left[ {\frac{1}{{P_{\mathop{\rm e}\nolimits} ^2}}{{\left( {\frac{{\partial {P_{\rm e}}}}{{\partial \varphi }}} \right)}^2} + \frac{1}{{P_{\mathop{\rm o}\nolimits} ^2}}{{\left( {\frac{{\partial {P_{\mathop{\rm o}\nolimits} }}}{{\partial \varphi }}} \right)}^2}} \right]^{ - \frac{1}{2}}}
\label{e10}
\end{equation}
with the probabilities of even and odd counts
\begin{align}
{P_{\mathop{\rm e}\nolimits} } &= \frac{1}{2}\left( {1 + \left\langle {\hat \Pi } \right\rangle } \right), 
\label{e11}\\ 
{P_{\mathop{\rm o}\nolimits} } &= \frac{1}{2}\left( {1 - \left\langle {\hat \Pi } \right\rangle } \right). 
\label{e12}
\end{align}
In terms of above calculation, we numerically obtain the optimal sensitivities with different mean photon numbers.
Figure \ref{f4} manifests the optimal sensitivity of our protocol, as a contrast, we also plot two reference curves that scale as $N^{-1}$ (Heisenberg limit) and $N^{-2}$.
The results reveal that a sub-Heisenberg-limited sensitivity is obtained since the optimal sensitivity lies in between $N^{-1}$ and $N^{-2}$ with $N \geqslant 2$.
For $N \leqslant 2$, the optimal sensitivity can even beat the $N^{-2}$.
This is an unreachable performance with respect to linear phase estimation using coherent and even exotic quantum states.

\section{Fundamental sensitivity limit}
\label{s4}
As mentioned previously, only through directly calculating the quantum Fisher information, one may deceive oneself into believing that there exists a superb sensitivity limit.
For example, if one uses operators $\exp \left[ {{{i\varphi \left( {{{\hat a}^\dag }\hat a - {b^\dag }b} \right)} \mathord{\left/
{\vphantom {{i\varphi \left( {{{\hat a}^\dag }\hat a - {b^\dag }b} \right)} 2}} \right.
\kern-\nulldelimiterspace} 2}} \right]$ and  $\exp \left( {i\varphi {{\hat a}^\dag }\hat a} \right)$, respectively, to describe a linear phase channel in a SU(2) interferometer with only a coherent state input, there are two different outcomes for optimal sensitivities deduced from quantum Fisher information, ${1 / {\sqrt N }}$ and ${1 / {\sqrt {2N} }}$, and the latter is inauthentic since it indicates a sub-shot-noise-limited sensitivity.

For keeping away from this pitfall, we invoke the phase-averaging approach to rule out any external resources that may provide phase information to the measurement device.
A simple understanding for the core of this approach is to disrupt the input state into a mixed state carrying random phase and losing all phase references.
After such treatment, an authentic fundamental sensitivity limit, without any hidden phase references, can be obtained via calculating quantum Fisher information, see Refs. \cite{PhysRevA.85.011801,PhysRevA.96.052118} for physical interpretation and detailed discussions.

According to the step of this framework, we consider the density matrix of the input state, 
\begin{equation}
{\rho _{{\rm{in}}}} = \sum\limits_{u= 0}^\infty \sum\limits_{v = 0}^\infty {{c_{uv}}\left| u \right\rangle \left\langle v \right| \otimes \left| 0 \right\rangle \left\langle 0 \right|}.
\label{e13}
\end{equation}
In the light of this density matrix, the phase-averaged input defined in phase-averaging approach is given by
\begin{align}
\nonumber \bar \rho  &= \frac{1}{{2\pi }}\int_0^{2\pi } {{{\hat V}_A}{{\hat V}_B}{\rho _{{\rm{in}}}}\hat V_A^\dag \hat V_B^\dag d\varphi }  \\ 
\nonumber &= \sum\limits_{u = 0}^\infty \sum\limits_{v = 0}^\infty {\int_0^{2\pi } {\frac{{d\varphi }}{{2\pi }}{e^{i\varphi \left( {u - v} \right)}}} {c_{uv}}\left| u \right\rangle \left\langle v \right| \otimes \left| 0 \right\rangle \left\langle 0 \right|}  \\ 
&= \sum\limits_{u = 0}^\infty  {{p_u}\left| u \right\rangle \left\langle u \right| \otimes \left| 0 \right\rangle \left\langle 0 \right|}.  
\label{e14}
\end{align}
Where ${{\hat V}_A} = \exp \left( {i\varphi {{\hat a}^\dag }\hat a} \right)$ and ${{\hat V}_B} = \exp \left( {i\varphi {{\hat b}^\dag }\hat b} \right)$, the probability $p_u = c_{uu}$ obeys $\sum\nolimits_{u = 0}^\infty  {{p_u} = 1}$.
Note that the phase-averaged input turns into a mixed state, and the external phase information is erased.
Then this input is incident on the first beam splitter, and the output density matrix has the following form
\begin{equation}
{\rho _1} = {{\hat U}_{\rm BS}}\bar \rho \hat U_{\rm BS}^\dag  = \sum\limits_{u = 0}^\infty  {{p_u}\left| {{\psi _u}} \right\rangle \left\langle {{\psi _u}} \right|}
\label{e15} 
\end{equation}
with
\begin{equation}
\left| {{\psi _u}} \right\rangle  = \sum\limits_{j = 0}^u {\sqrt {\frac{{u!}}{{j!\left( {u - j} \right)!}}} } {\left( {\frac{1}{{\sqrt 2 }}} \right)^u}\left| j \right\rangle \left| {u - j} \right\rangle.
\label{e16}
\end{equation}
By using the convexity of the quantum Fisher information \cite{PhysRevA.63.042304} and the orthogonality of the states $\left\langle {{{\psi _{u'}}}}
{\left | {\vphantom {{{\psi _{u'}}} {{\psi _u}}}}
		\right. \kern-\nulldelimiterspace}
	{{{\psi _u}}} \right\rangle  = {\delta _{uu'}}$, we have
\begin{equation}
{\cal F}_{\rm q} = \sum\limits_{u = 0}^\infty  {{p_u}{\cal F}_{\rm q}^u},
\label{e17}
\end{equation}
The quantum Fisher information for each orthogonal component is
\begin{equation}
{\cal F}_{\rm q}^u = 4\left[ {\left\langle {{{\left( {{{\hat a}^\dag }a} \right)}^4}} \right\rangle  - {{\left\langle {{{\left( {{{\hat a}^\dag }a} \right)}^2}} \right\rangle }^2}} \right],
	\label{e18}
\end{equation}
where the expectation values are taken over the state $\left| {{\psi _u}} \right\rangle$.

\begin{figure}[htbp]
\centering\includegraphics[width=7cm]{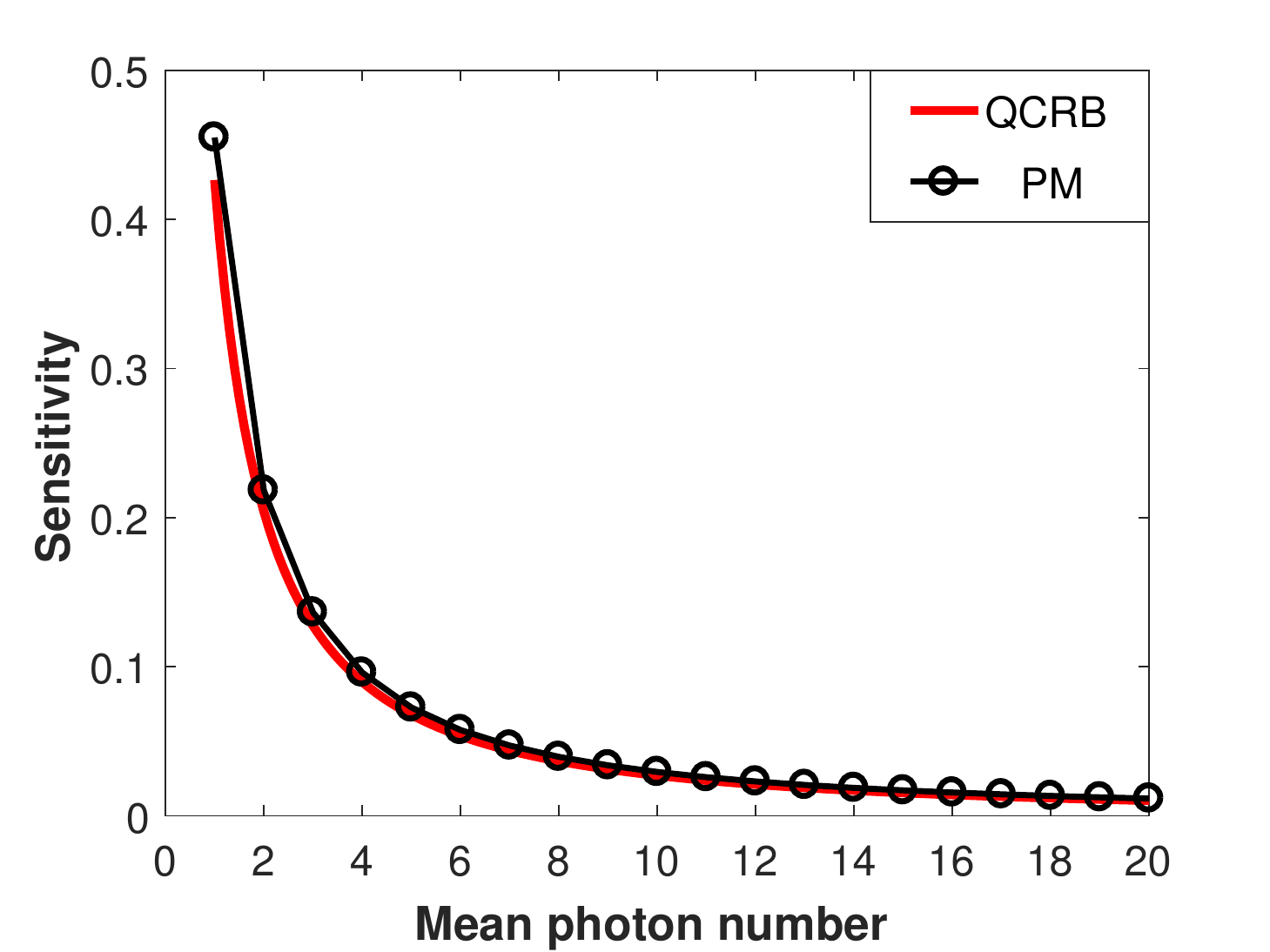}
\caption{The optimal sensitivity and quantum Cram\'er-Rao bound with parity measurement as functions of the mean photon number.}
\label{f5}
\end{figure}

With the boson algebra $\left[ {\hat a,{{\hat a}^\dag }} \right] = 1$, one can get the normal order form of the operators,
\begin{align}
{\left( {{{\hat a}^\dag }a} \right)^4} &= {{\hat a}^{\dag 4}}{{\hat a}^4} + 6{{\hat a}^{\dag 3}}{{\hat a}^3} + 7{{\hat a}^{\dag 2}}{{\hat a}^2} + {{\hat a}^\dag }\hat a, 
\label{e19}\\ 
{\left( {{{\hat a}^\dag }a} \right)^2} &= {{\hat a}^{\dag 2}}{{\hat a}^2} + {{\hat a}^\dag }\hat a.
\label{e20} 
\end{align}
By means of the summation formula in Ref. \cite{PhysRevA.96.052118}, the expectation values in Eqs. (\ref{e19}) and (\ref{e20}) can be calculated,
\begin{align}
\left\langle {{{\hat a}^{\dag 4}}{{\hat a}^4}} \right\rangle  &= \frac{1}{{{2^4}}}u\left( {u - 1} \right)\left( {u - 2} \right)\left( {u - 3} \right), 
\label{e21}\\ 
\left\langle {{{\hat a}^{\dag 3}}{{\hat a}^3}} \right\rangle  &= \frac{1}{{{2^3}}}u\left( {u - 1} \right)\left( {u - 2} \right), 
\label{e22}\\ 
\left\langle {{{\hat a}^{\dag 2}}{{\hat a}^2}} \right\rangle & = \frac{1}{{{2^2}}}u\left( {u - 1} \right), 
\label{e23}\\ 
\left\langle {{{\hat a}^\dag }\hat a} \right\rangle  &= \frac{1}{2}u. 
\label{e24}
\end{align}
Combining Eqs. (\ref{e18})-(\ref{e24}), we get
\begin{equation}
{\cal F}_{\rm q}^u = \frac{u}{2}\left( {u + 1} \right)\left( {2u - 1} \right).
\label{e25}
\end{equation}
Further, the quantum Fisher information corresponding to fundamental sensitivity limit thus reads
\begin{equation}
{{\cal F}_{\rm q}} = {e^{ - N}}\sum\limits_{u = 0}^\infty  {\frac{{{N^{u - 1}}}}{{2\left( {u - 1} \right)!}}\left( {u + 1} \right)\left( {2u - 1} \right)}.
	\label{e26}
\end{equation}
As a criterion to appraise our strategy, in Fig. \ref{f5} we plot the sensitivity with parity measurement, and the quantum Cram\'er-Rao bound associated with quantum Fisher information via 
$\delta {\varphi _{\rm QCRB}} = {1 \mathord{\left/
			{\vphantom {1 {\sqrt {{\cal F_{\rm q}}} }}} \right.
			\kern-\nulldelimiterspace} {\sqrt {{\cal F_{\rm q}}} }}$.
It can be seen that our strategy approximately overlap with the bound, i.e., parity measurement is a quasi-optimal strategy for our protocol.
	
\section{Effects of realistic scenarios}
\label{s5}
For a practical estimation system, the disturbance is inevitable.
In this section we discuss the effects arising from several realistic scenarios consisting of photon loss, imperfect detector, and those which are a combination thereof.
	
\subsection{Effect of photon loss}
It is well known that no metrological systems can thoroughly circumvent the influence originating from the environment, known as quantum decoherence, for the systems themselves are immersed in the environment.
To such decoherence process there corresponds to the information leakage from the system to the environment, and then, the performance of the system will degenerate.
Hence, for more practical to our protocol, in this section we discuss the effect of photon loss on the resolution and the sensitivity.
	
In an optical propagation process, loss is inevitable as photons will be partially absorbed by environment.
For this scenario, the relationship between a lossy channel and a single-mode quantum state $\rho$ is generally delineated by a completely positive map ${{\cal S}_\eta }$ of the form \cite{Kraus1983States},
\begin{equation}
{{\cal S}_\eta }\left[ \rho  \right] = \sum\limits_{l = 0}^\infty  {{{\hat K}_{\eta ,l}}\rho \hat K_{\eta ,l}^\dag }.
\label{e27} 
\end{equation}
Where ${{\cal S}_\eta }$ is the so-called super-operator, and
\begin{equation}
{\hat K_{\eta ,l}} = \frac{{{{\left( {1 - \eta } \right)}^{\frac{l}{2}}}{\eta ^{\frac{{\hat n}}{2}}}{{\hat a}^l}}}{{\sqrt {l!} }}
\label{e28}
\end{equation}
is Kraus operator, the ratio of photon loss holds true for $ {1 - \eta } =L \in \left[ {0,1} \right]$. 
	
Indeed, the lossy channel is usually simulated by inserting two linear beam splitters with transmissivities $\eta_A$ and $\eta_B$, the reflected photons that enter into the environment are viewed as photon loss \cite{Gard2017,PhysRevA.82.053804}.
In the following discussion, we assume that the loss occurs before the nonlinear channel.
	
On account of the aforementioned discussions, in the presence of photon loss, the expectation value of parity operator goes to
\begin{widetext}
\begin{align}
\nonumber {\left\langle {\hat \Pi } \right\rangle _{\rm R1}} &= {\rm{Tr}}\left\{ {{{\cal S}_{{\eta _A}}}\left[ {{\rho _A}} \right]{{\cal S}_{{\eta _B}}}\left[ {{\rho _B}} \right]\left[ {{{\hat I}_A} \otimes \exp \left( {i\pi {{\hat n}_B}} \right)} \right]} \right\}\\
&=\exp \left[ { - \frac{N}{2}\left( {{\eta_A} + {\eta_B}} \right)} \right]\sum\limits_{n = 0}^\infty  {\sum\limits_{m = 0}^\infty  {\frac{{{{\left( {{{\sqrt {{\eta_A}{\eta_B}} N{e^{in\varphi }}} \mathord{\left/
									{\vphantom {{\sqrt {{\eta_A}{\eta_B}} N{e^{in\varphi }}} 2}} \right.
									\kern-\nulldelimiterspace} 2}} \right)}^n}}}{{n!}}} } \frac{{{{\left( {{{\sqrt {{\eta_A}{\eta_B}} N{e^{ - im\varphi }}} \mathord{\left/
								{\vphantom {{\sqrt {{\eta_A}{\eta_B}} N{e^{ - im\varphi }}} 2}} \right.
								\kern-\nulldelimiterspace} 2}} \right)}^m}}}{{m!}},
\label{e29}
\end{align}
\end{widetext}
further, the sensitivity is calculated via the classical Fisher information.
	
\begin{figure}[htbp]
\centering\includegraphics[width=7cm]{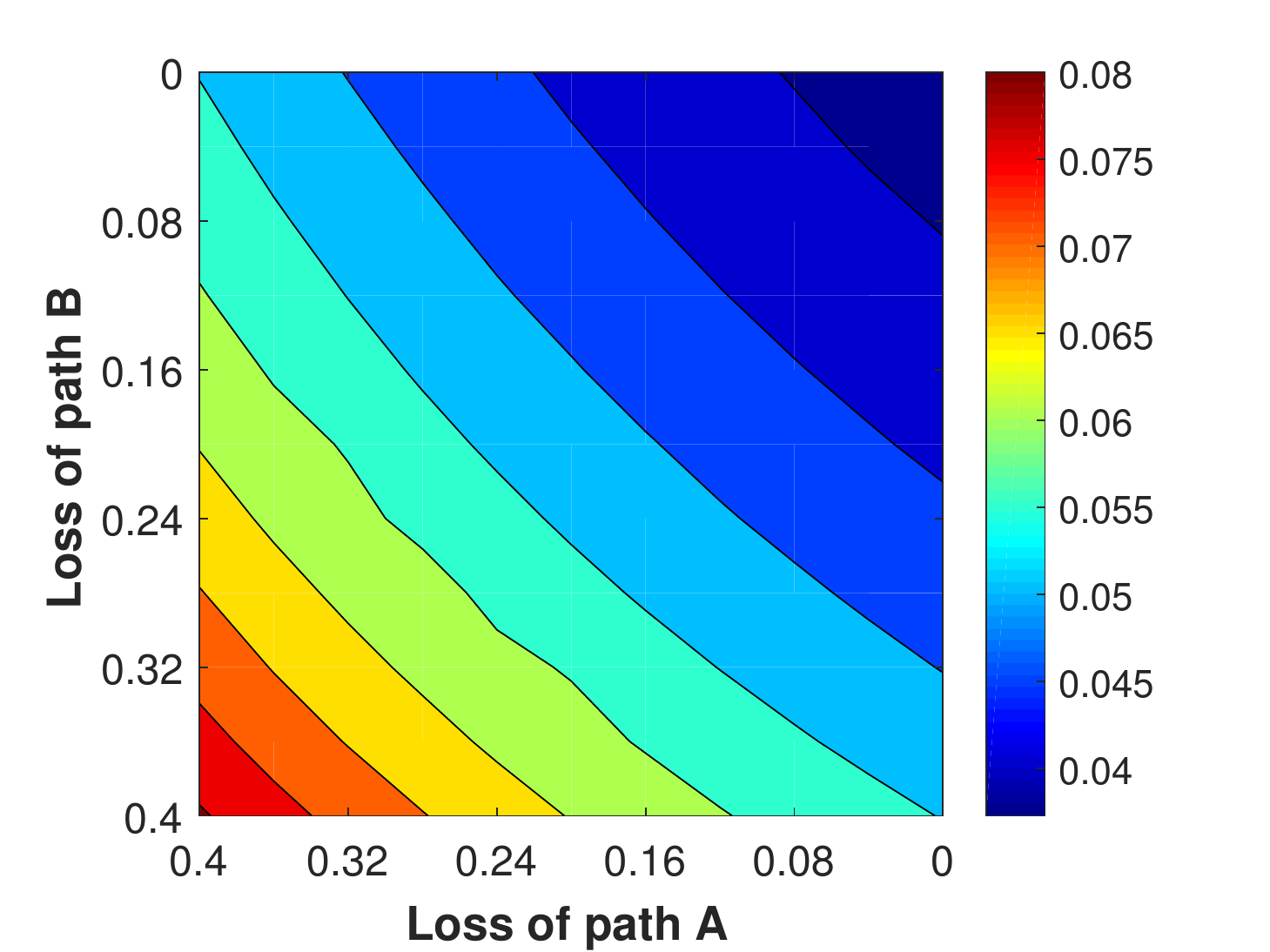}
\centering\includegraphics[width=7cm]{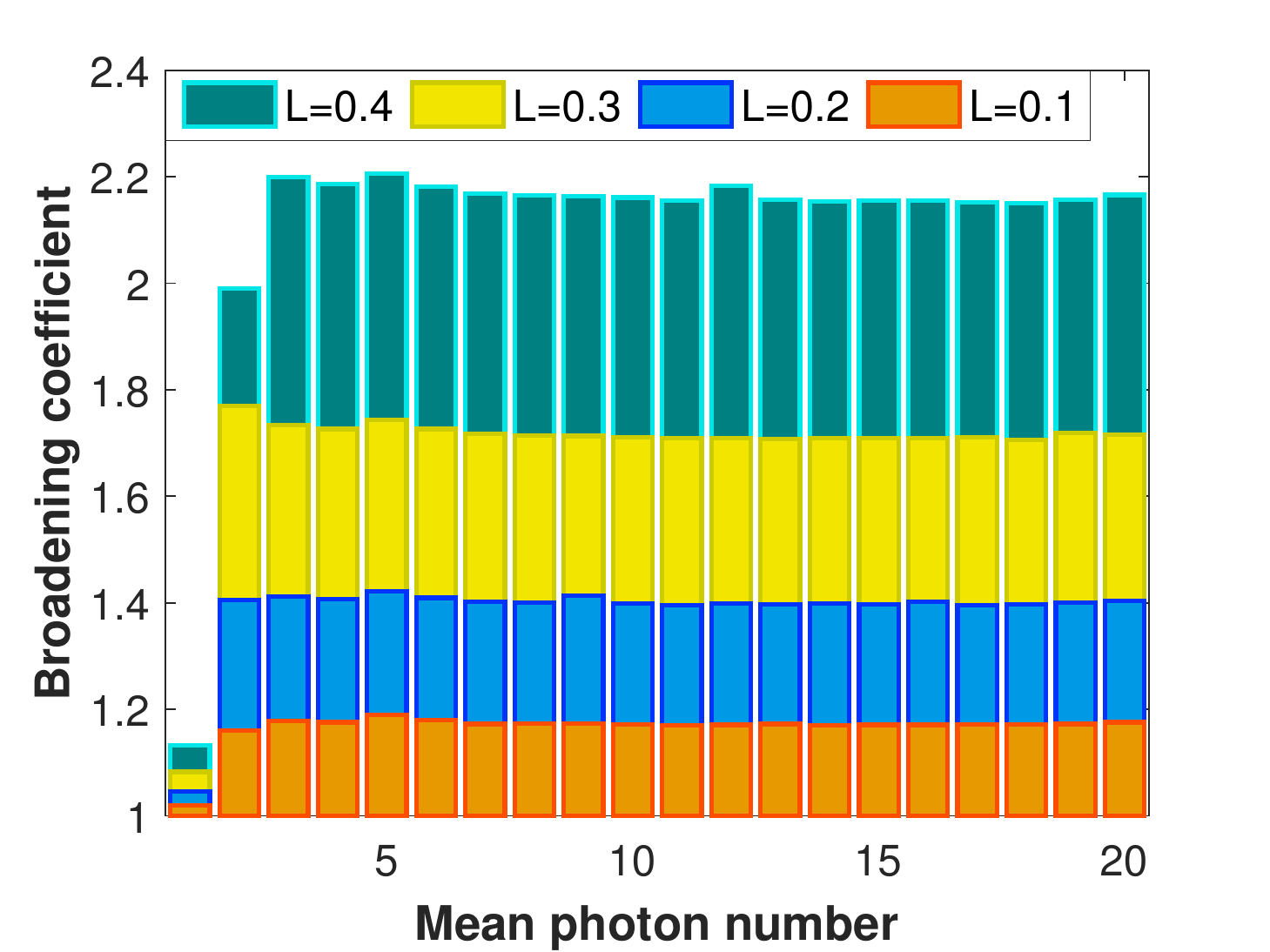}
\caption{(a) The FWHM with parity measurement under the photon loss as a function of the losses of two paths in the interferometer, where $N=10$. The color bar on the right of the 2D-planar forms indicates the corresponding value.
(b) The broadening coefficient as a function of mean photon number in the case of $L_A=L_B=L$, where $L$ takes on the values of 0.4, 0.3, 0.2 and 0.1.}
\label{f6}
\end{figure}

To investigate the effect of loss on the resolution, we plot FWHMs with different losses of two paths, where $N=10$.
Figure \ref{f6}(a) suggests that the FWHM of identical two losses is slightly superior to that of unequal two losses, because the contour lines are marginally convex compared with diagonal lines, $L_A+L_B={\rm constants}$.
In addition, the minima of signals with different losses sit at 0, in turn, the visibility maintains 100\% whatever photon loss is.  
With $L_A=L_B=L$, we study the deterioration of resolution associated with the broadening of FWHM, while the broadening coefficient is defined as
    
\begin{equation}	
{C_{\mathop{\rm B}\nolimits} } = {\left. {\frac{{{\mathop{\rm FWHM}\nolimits} \left[ {{{\left\langle {\hat \Pi } \right\rangle }_{\rm R1}}} \right]}}{{{\mathop{\rm FWHM}\nolimits} \left[ {\left\langle {\hat \Pi } \right\rangle } \right]}}} \right|_L}.
\label{e30}
\end{equation}
	
Figure \ref{f6}(b) exhibits the broadening coefficients with different losses.
For $N \geqslant 3$ and any loss, the coefficient has a slight fluctuation near a fixed value as the increase of mean photon number.
This suggests that the broadening of FWHM is only related to loss and has nothing to do with mean photon number, the super-resolved characteristic is robust as $C_{\rm B}<2.3$ even at a total loss of 40\% ($L_A=L_B=0.4$).

\begin{figure}[htbp]
\centering\includegraphics[width=7cm]{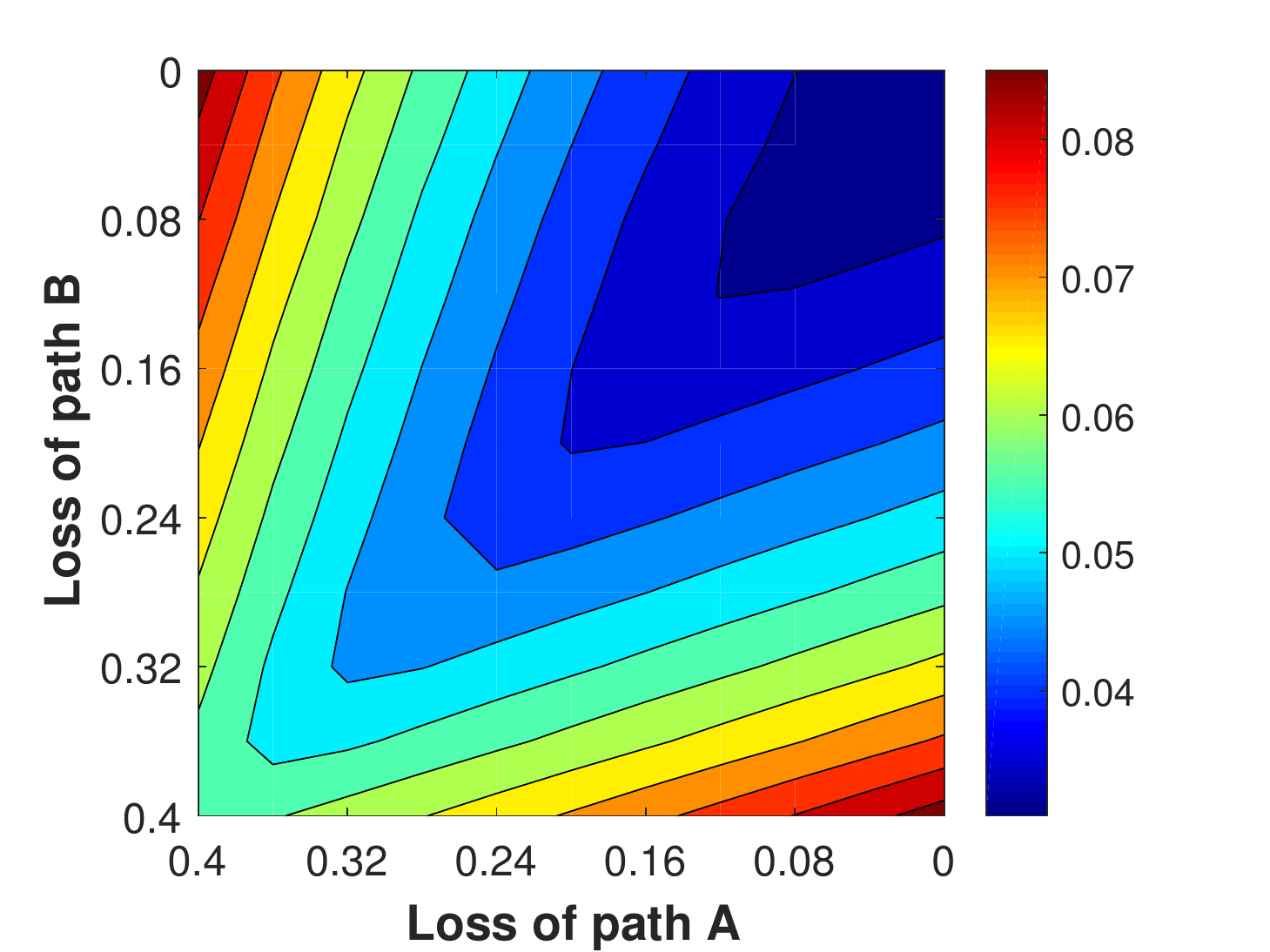}
\caption{The optimal sensitivity with parity measurement under the photon loss as a function of the losses of two paths in the interferometer, where $N=10$. The color bar on the right of the 2D-planar forms indicates the corresponding value.}
\label{f7}
\end{figure}

Figure \ref{f7} depicts the sensitivities for different losses with $N=10$.
It is obvious that all the sensitivities surpass the Heisenberg limit ($N^{-1}=0.1$) even if the total loss is 40\%, and this reveals that our measurement strategy is robust.  
Another interesting phenomenon is that once two losses are unequal, the sensitivity gets worse no matter how we suppress the loss in either of the two paths, e.g., the sensitivity with $L_A = L_B=0.4$ is better than that with $L_A < L_B  = 0.4$.
An understanding to this result, two same intensities in the two paths maintain the optimal indistinguishability, however, unequal intensities partially erase it and then destroy the sensitivity.
As a result, one can find that, under the identical total loss, the sensitivity is better when the losses of the two paths are adjacent, e.g., the sensitivities with diagonal line $L_A+L_B =0.4$ in the figure.

\subsection{Effect of imperfect detector}
Apart from the disturbance arising out of environment, there exists other realistic scenarios, dark counts and response-time delay, which stem from imperfect detector. 
For parity measurement, a photon-number-resolving detector is often used to record the parity of each measuring trial \cite{Cohen:14,Lundeen2008Tomography}, and the dark counts will disturb the parity. 
Suppose that the rate of dark counts is $r$, at either of two outputs, the probability of $w$ dark counts follows the Poissonian distribution
		
\begin{equation}
{P_{\rm d}}\left( w \right) = {e^{ - r}}\frac{{{r^w}}}{{w!}}.
\label{e31}
\end{equation}
Obviously, even dark counts has no effect on the parity.
Thus we rewrite the probability of odd counts in the realistic measurement as
	
\begin{align}
\nonumber {P'_{\rm o}} &= {P_{\rm o}}\sum\limits_{w = \rm even} {{P_{\rm d}}\left( w \right)}  + {P_{\rm e}}\sum\limits_{w = \rm odd} {{P_{\rm d}}\left( w \right)} \\
&= \frac{1}{2}\left( {1 - {e^{ - 2r}}\left\langle {\hat \Pi } \right\rangle } \right),
\label{e32}
\end{align}
and the probability of even counts is equal to
\begin{equation}
{P'_{\rm e}} = \frac{1}{2}\left( {1 + {e^{ - 2r}}\left\langle {\hat \Pi } \right\rangle } \right).
\label{e33}
\end{equation}
One may easily derive the expectation value of parity operator as
\begin{equation}
{\left\langle {\hat \Pi } \right\rangle _{\rm R2}} = {e^{ - 2r}}\left\langle {\hat \Pi } \right\rangle .
\label{e34}
\end{equation}

Response-time delay is the other realistic problem in the detector.
Here we offer the interpretation about the relationship between the response-time delay and dark counts.
In the practical measurement, it is not a fixed value for the response-time, and the time delay can be expressed as a mean time delay along with a delay jitter $\tau$. 
Thereinto, the delay of the mean value has no effect on the measuring outcomes, for the strategy in this paper is to count the photon number rather than arriving time. 
The delay jitter, nevertheless, causes an increase in the width of the sampling detection gate.

\begin{figure}[htbp]
\centering
\includegraphics[width=7.5cm]{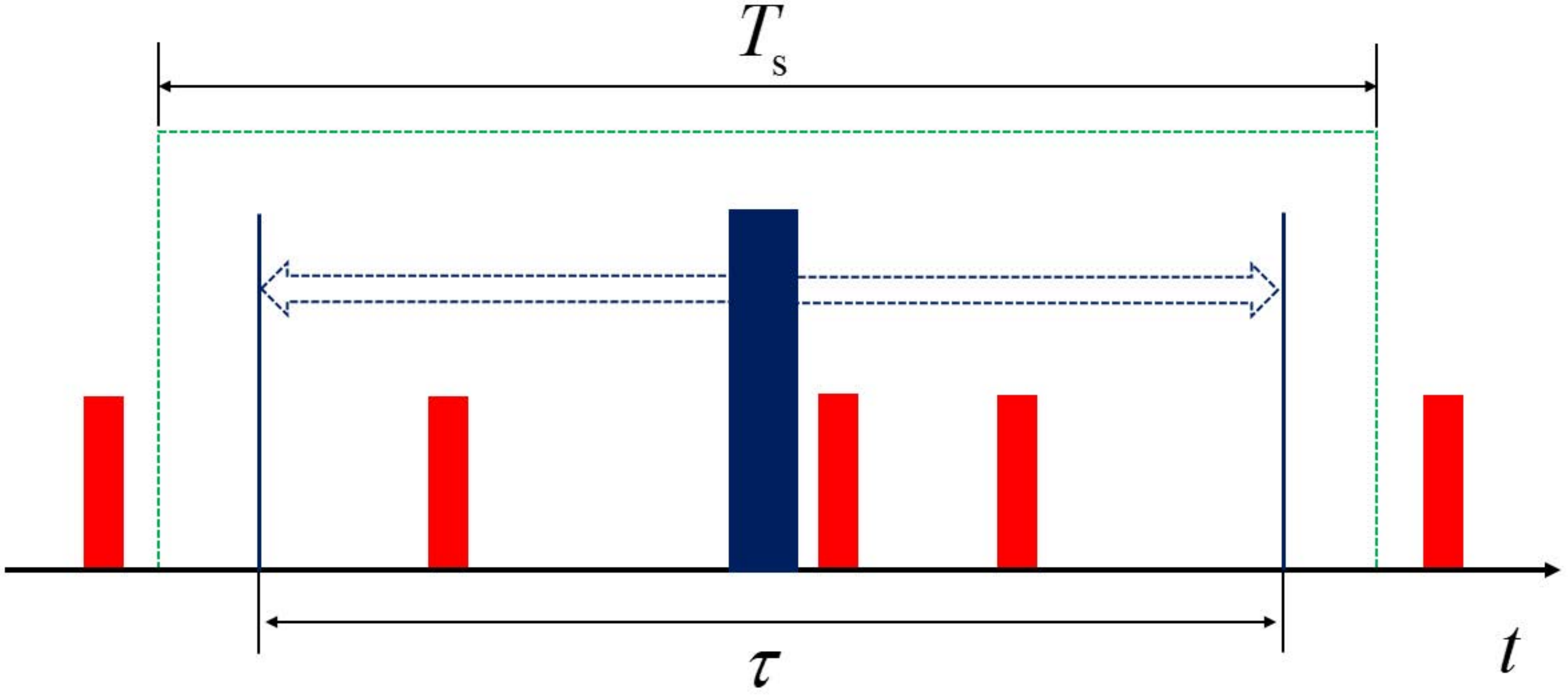}
\caption{Schematic of the effect of response-time delay on the measuring outcomes.}
\label{f8}
\end{figure}

Schematic diagram for the effect of the response-time delay on the measuring outcomes is shown in Fig. \ref{f8}, where the blue rectangle is the theoretical standard response-time. 
The time of the actual arriving signal may occur in a time period $\tau$ due to the presence of the response time delay. 
$T_{\rm s}$ is the time width of sampling detection gate and the relationship $\tau \leqslant T_{\rm s}$ is satisfied to guarantee merely single trigger in each gate. 
The red rectangle implies the pulse of dark counts and its distribution is random, however, the statistical results obey the Poissonian distribution. 
Moreover, the dark counts outside the sampling detection gate do not affect the measurement. 
The width of the gate has to be increased owing to the response-time delay, hence, the direct influence of time delay on the measuring outcomes is to increase the rate of dark counts.

In general, the width increase of sampling gate is less than an order of magnitude, we describe the joint effect of dark counts and response-time delay by $d=10r$. 
At this time, the Eq. (\ref{e34}) may be revised as

\begin{equation}
{\left\langle {\hat \Pi } \right\rangle _{\rm R2}} = {e^{ - 2d}}{\left\langle {\hat \Pi } \right\rangle},
\label{e35}
\end{equation}

Equation (\ref{e35}) points out that the signal is modulated by a factor ${e^{ - 2d}}$.
This factor will lead to the whole decline in signal amplitude, however, the FWHM and visibility of the signal are changeless owing to the fact that the signal minimum is 0. 

\begin{figure}[htbp]
\centering
\includegraphics[width=7cm]{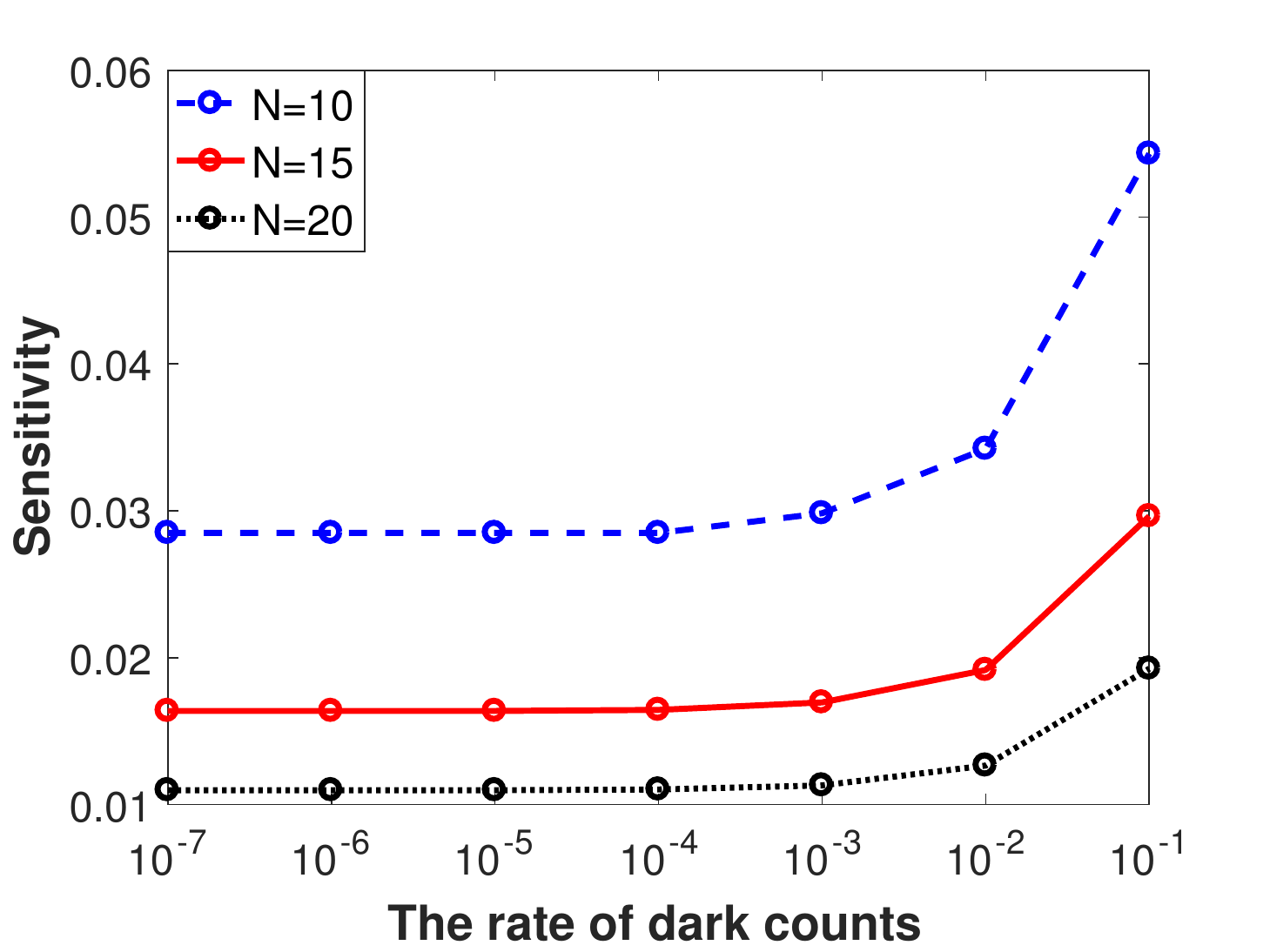}
\caption{The optimal sensitivity as a function of the rate of dark counts, where $N$ takes on the values of 10, 15 and 20.}
\label{f9}
\end{figure}

Under the current technical conditions, the rate of dark counts $r$ in a commercial detector ranges from $10^{-8}$ to $10^{-2}$ \cite{Hadfield2007Single}, i.e.,  the rate $d$ pertains to the interval $10^{-7}\leqslant d \leqslant 10^{-1}$.
In Fig. \ref{f9}, we plot the optimal sensitivity as a function of the rate of dark counts.
The results show that, for the region of $d \geqslant 10^{-3}$, the sensitivity gets worse as increasing the rate of dark counts,  and the sensitivity can be considered as unaffected with $d  < 10^{-3}$.

\subsection{Joint effect of photon loss and imperfect detector }
For the realistic measurements, neither photon loss nor imperfect detector can be ignored, and in general, the former and the latter will exist simultaneously.	
In this section, we investigate the effect in the presence of photon loss and imperfect detector simultaneously.
On account of the conclusions in the above sections, the expectation value of parity operator in such scenario is expressed as 	
	
\begin{equation}
{\left\langle {\hat \Pi } \right\rangle _{\rm R3}} = {e^{ - 2d}}{\left\langle {\hat \Pi } \right\rangle _{\rm R1}},
\label{e36}
\end{equation}
where the expectation value ${\left\langle {\hat \Pi } \right\rangle _{\rm R1}}$ has been defined in Eq. (\ref{e29}), and the sensitivity can be obtained by classical Fisher information.

\begin{figure}[htbp]
	\centering
	\includegraphics[width=7cm]{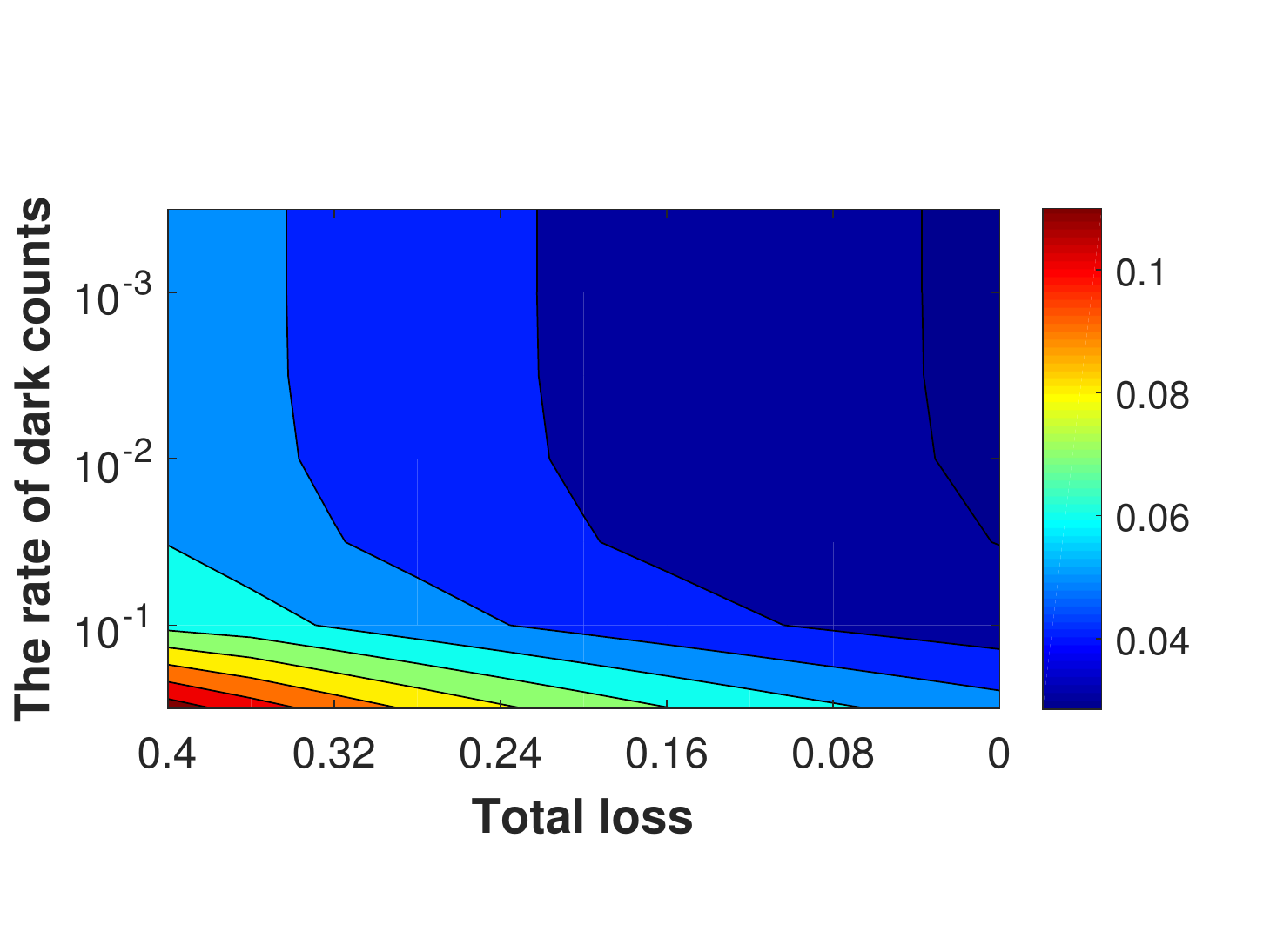}
	\caption{The optimal sensitivity as a function of both total loss and the rate of dark counts, where $N=10$. The color bar on the right of the 2D-planar form indicates the corresponding value.}
	\label{f10}
\end{figure}	
	
This scenario is the same as the case in the previous section, the factor ${e^{ - 2d}}$ has no effect on both the FWHM and visibility of signal.
With regard to the sensitivity, in Fig. \ref{f10}, we demonstrate the optimal sensitivity as a function of both total loss and the rate of dark counts. 
In the presence of low loss, the degeneration rooted in dark counts on the sensitivity is inconspicuous.
On the other hand, for high lossy region, only if the rate of dark counts satisfies $d<10^{-2}$ does the sensitivity keep constant approximately.
The severely affected sensitivities appear at the region of simultaneously high loss and dark counts, and the sensitivity at this time can not achieve the Heisenberg limit.
Therefore, keeping low either loss or dark counts is an effective way to achieve the sub-Heisenberg limit.

\section{Conclusion}
\label{s6}
In summary, we present a nonlinear phase estimation protocol.
A coherent state and parity measurement play the roles of input and detection strategy, respectively.
The parity signal pertaining to our protocol has super-resolved characteristic boosted by a factor of $N$ or so compared with the signal of linear phase estimation, meanwhile, the signal possesses a 100\% visibility.
On the other hand, we obtain a sub-Heisenberg-limited sensitivity, which lies in between $N^{-1}$ and $N^{-2}$.
Additionally, using phase-averaging approach to eliminate any hidden resources, we ascertain the low-down on the fundamental sensitivity limit derived from quantum Fisher information, and find that our strategy exceedingly approaches this limit.

Finally, for more practical in applications, we discuss the effects of several realistic scenarios on the resolution and the sensitivity, including photon loss, imperfect detector, and their combination.
The results reveal that identical losses are propitious to maintain favorable performances, both resolution and sensitivity.
Meanwhile, the sensitivity holds sub-Heisenberg limit and the FWHM has a 2.2 times broadening even if there exists 40\% total loss. 
The imperfect detector has the apparent effects on the sensitivity only in the region of $d \geqslant 10^{-3}$, whereas neither the resolution nor the visibility are affected by the detector.
When the two scenarios exist simultaneously in the measurement, a sub-Heisenberg-limited sensitivity is reachable so long as either photon loss or dark counts is low.
Overall, our protocol achieves super-resolved and super-sensitive estimation for nonlinear phase shift, and parity measurement is a quasi-optimal and robust strategy.

\section*{Funding}
National Natural Science Foundation of China (Grant No. 61701139).

\end{document}